\documentclass[useAMS]{mn2e}

\usepackage{graphicx}                                                                                                                                         
\usepackage{amssymb}
\usepackage{rotating}

\title[Phase-resolved UV spectroscopy of HD~191612]{Phase-resolved ultraviolet spectroscopy of the magnetic Of?p star HD~191612}

\author[W. L. F. Marcolino et al.]
{\parbox{\textwidth}{W. L. F. Marcolino$^{1}$,\thanks{E-Mail: wagner@astro.ufrj.br}
J.-C. Bouret$^{2,3}$, 
J. O. Sundqvist$^{4}$,
N. R. Walborn$^{5}$,
A. W. Fullerton$^{5}$,
I. D. Howarth$^{6}$,
G. A. Wade$^{7}$ and 
A. ud-Doula$^{8}$}\vspace{0.4cm}\\
\parbox{\textwidth}{$^{1}$Observat\'orio do Valongo, Universidade Federal do Rio de Janeiro, Ladeira Pedro Ant\^onio, 43, CEP 20080-090, Brasil \\
$^{2}$LAM-UMR6110, CNRS \& Universit\'e Provence, rue Fr\'ederic Joliot-Curie, F-13388 Marseille Cedex 13, France    \\
$^{3}$NASA/GSFC, Code 665, Greenbelt, MD 20771, USA\\
$^{4}$University of Delaware, Bartol Research Institute, Newark, Delaware 19716, USA\\
$^{5}$Space Telescope Science Institute, 3700 San Martin Drive, Baltimore, MD 21218, USA\\
$^{6}$Department of Physics and Astronomy, University College London, Gower Street, London, WC1E 6BT, UK\\
$^{7}$Department of Physics, Royal Military College of Canada, PO Box 17000, Station Forces, Kingston, ON K7K 7B4, Canada\\ 
$^{8}$Penn State Worthington Scranton, 120 Ridge View Drive, Dunmore, PA 18512, USA\\
}}

\begin{document}   

\date{Received; Accepted }

\pagerange{\pageref{firstpage}--\pageref{lastpage}} \pubyear{2012}

\maketitle
\label{firstpage}

\begin{abstract}

We present for the first time phase-resolved UV spectroscopy of an Of?p star, namely, 
HD~191612. The observations were acquired with the Space Telescope Imaging 
Spectrograph ({\em STIS}) on-board the Hubble Space Telescope ({\em HST}). We report the 
variability observed in the main photospheric and wind features and compare the results with 
previous findings for the Of?p star HD~108. We show that UV line strengths, H$\alpha$, and  
longitudinal magnetic field, vary coherently according to the rotational period (P$_{rot}$ = 537.6d), 
providing additional support for the magnetic oblique rotator scenario. The stellar and wind 
parameters of HD~191612 are obtained based on NLTE expanding atmosphere models. The peculiar wind line 
profile variations revealed by the new {\em STIS} data - not reproduced by 1D atmosphere models - are addressed through non-spherical 
MHD simulations coupled with radiative transfer. The basic aspects of the UV variability observed 
are explained and the structure of the {\it dynamical magnetosphere} of HD~191612 is discussed.

\end{abstract}



\maketitle


\section{INTRODUCTION}

The peculiar Of?p stars have been studied in detail in the
optical part of the spectrum, with observations spanning several
decades for HD~108 and HD~191612 (e.g., Walborn 1973; Naz\'e et al. 2010). 
Large-amplitude variability of optical features (e.g., of H$\alpha$ and He I lines) 
is found to be common for this spectral class (Naz\'e, Vreux \& Rauw 2001; 
Howarth et al. 2007), and is characterized by `high' and `low' (or maximum/minimum) 
states in the degree of emission.  This behaviour is now understood in the context 
of a magnetic oblique rotator model, wherein the observer's view of a 
magnetically constrained stellar-wind outflow is modulated by stellar rotation 
(see Donati et al. 2006; Wade et al. 2011; Sundqvist et al. 2012).

Despite advances in understanding the nature of Of?p stars,
the behavior of the ultra\-violet (UV) spectra at different optical states
remains poorly known. Such information is crucially important, since this spectral
region contains the most sensitive diagnostics of hot-star winds.  We
therefore lack, so far, a complete view of the stellar-wind
confinement phenomenon (ud-Doula \& Owocki 2002) in these stars. A first
effort to address this problem was recently reported by our group for
the very slow rotator HD~108 (Marcolino et al. 2012; hereinafter
Paper~I). New observations in the ultra\-violet were acquired
close to the optical minimum state; the UV spectrum did show changes
compared to International Ultraviolet Explorer ({\em IUE}) observations acquired 
during a high state, but the variability was not as dramatic as observed in the optical.  
Marcolino et al. (2012) concluded that the stellar wind of HD~108 is not severely
affected by the magnetic field on a global scale (i.e., at large radii; 
see Paper~I for details). However, because of the long
rotational period of HD~108 (probably about 55 years; Naz\'e, Vreux \& Rauw 2001; Naz\'e et al. 2010), 
we still lacked information about UV line profiles at various rotational phases. 
Thus, until now, we have not had a detailed knowledge of UV variability matching 
that in the optical, for any Of?p star.

In this paper we present high-resolution UV spectroscopy of 
HD~191612\footnote{HD~191612 is a known double-lined spectroscopic binary, with $P_{orb} = 1542$d (Howarth et al. 2007).}
($P_{\rm rot} = 537.6$~d; Howarth et al. 2007), which represents the first
phase-resolved UV spectra of an Of?p star covering the entire
rotational cycle. The observations are described in
Section~\ref{obs}, where we also summarize the variability observed in the main
photospheric and wind features and compare the results with our previous findings 
for HD~108. We demonstrate that the strengths of the main UV wind lines, H$\alpha$, 
and longitudinal magnetic field ($B_Z$), vary coherently according to the rotational 
period, thereby providing support for an origin linked to the magnetic oblique
rotator model. The stellar and wind parameters of HD~191612 are then
revised and discussed based on UV modelling using non-LTE expanding-atmosphere 
models (Section~\ref{models}). The peculiar variability revealed by the 
new {\em STIS} data is not reproduced by 1D models, but can be explained qualitatively by using 
magnetohydrodynamical (MHD) simulations of a magnetically confined wind (Section \ref{mhd}). 
Section~\ref{concs} summarizes the main results of our work and discuss 
their consequences for the Of?p class as a whole.
 
\begin{figure*}                          
\includegraphics[width=16cm,height=21cm,angle=-180]{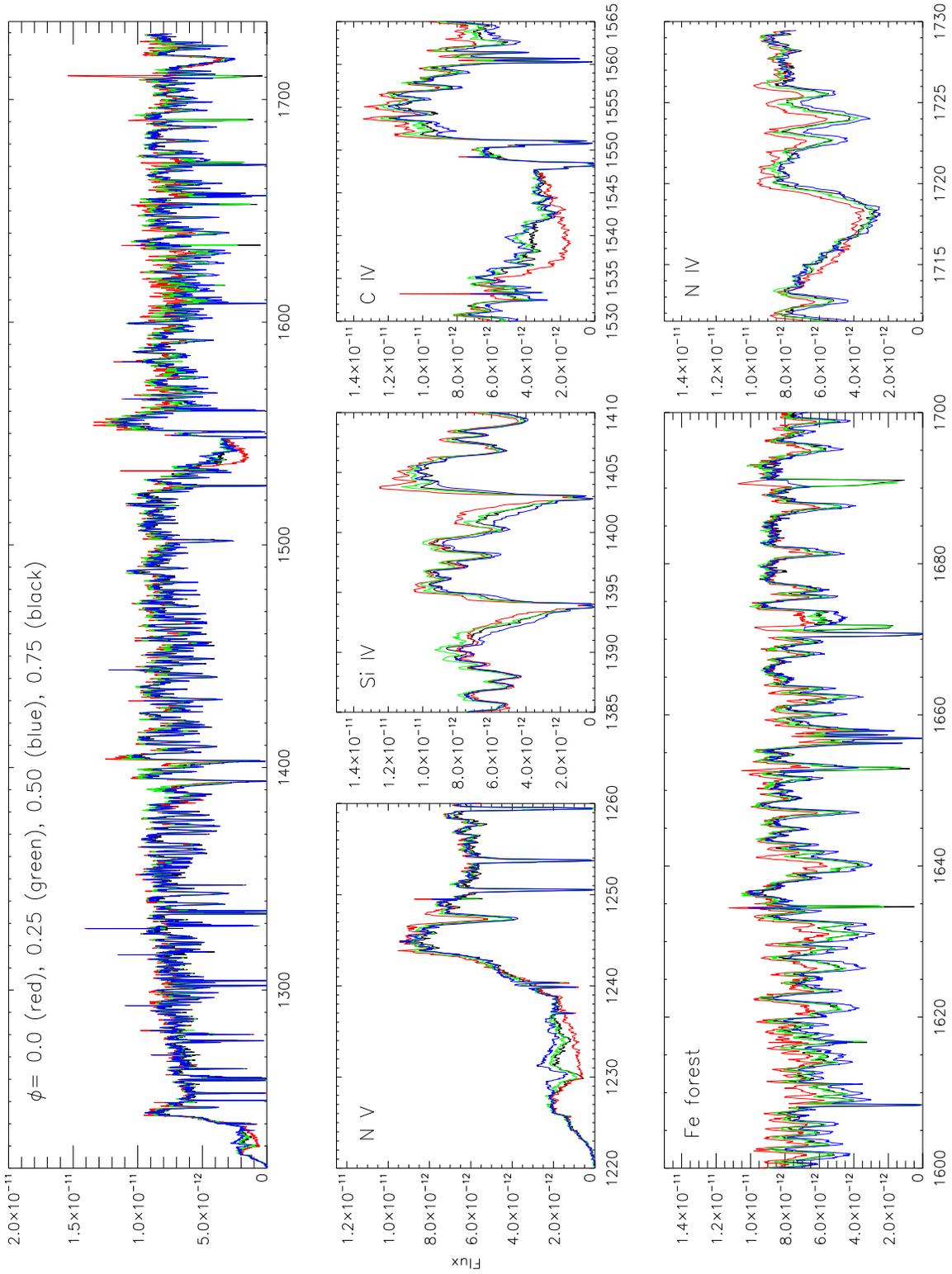}       
\caption{Phase-resolved data of HD~191612: $\phi$ = 0.0, 0.25, 0.50, and 0.75. The continua are essentially the same, the total energy 
emitted in the UV is essentially preserved. The N$\;${\sc v} and C$\;${\sc iv} profile changes from $\phi$ = 0.0 and 0.5 follow the trend observed in 
HD~108 (high to low state; Marcolino et al. 2012). Flux units are erg cm$^{-2}$ s$^{-1}$ \AA$^{-1}$. Wavelengths are given in angstroms.}      
\label{uv}       
\end{figure*}


\section{OBSERVATIONS}
\label{obs}

We used the Space Telescope Imaging Spectrograph ({\em STIS}) on board {\em HST} to obtain 
ultraviolet spectra of HD 191612 (PI. Bouret; program ID 12179). Table \ref{obsdetails} provides 
a log of the observations, which were obtained at four phases separated by a 
quarter of the star's 537.6d rotational cycle. At each phase, the observations consisted of two pairs of exposures 
through the small ($0.2" \times 0.2"$) aperture, first with the E140M echelle grating 
and then with the E230M echelle grating. 

Observations with the E140M grating were centered at 1425~{\AA} and recorded by the FUV MAMA detector,
while observations with the E230M grating were centered at 2707~{\AA} and recorded by the NUV MAMA detector.
The total wavelength coverage is $\sim$1150 -- 1750~\AA\, (E140M) and $\sim$2250 -- 3100~\AA\, 
(E230M), with resolving powers R = 45,800 and 30,000, respectively. 
The modified Julian Dates (MJD) of the mid-exposures are listed in Table \ref{obsdetails}, 
along with the total integration time (divided equally between the two exposures). Since the two exposures 
obtained with each configuration were combined, the mid-point of the sequence was 
used to calculate the phase $\phi$ according to the ephemeris published by 
Howarth et al. (2007). Due to the very long stellar variability period, 
the NUV/FUV exposure pairs correspond to essentially identical phases 
(to within a few 10$^{-5}$ cycles).


\subsection{Observed variability characteristics} 
\label{variab}

\begin{figure}       
\centering       
\includegraphics[width=4.7cm,height=8.5cm,angle=90]{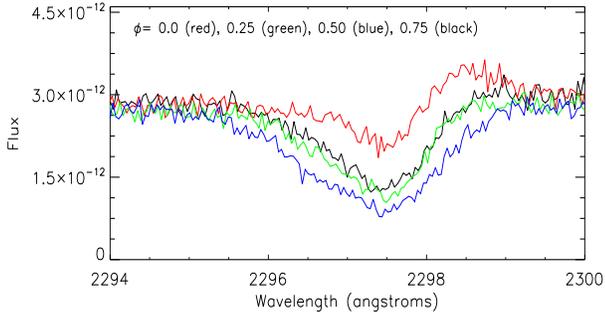}       
\caption{The C$\;${\sc iii}~$\lambda$2296 feature at different rotational phases - NUV/MAMA {\em STIS} spectra. 
Flux units are erg cm$^{-2}$ s$^{-1}$ \AA$^{-1}$.}      
\label{nuv}       
\end{figure}

In Figure \ref{uv} we present the spectra of HD~191612 obtained at different 
rotational phases. In this section we focus mainly on FUV data, where we find the main stellar 
wind diagnostics (N$\;${\sc v}~$\lambda$1240, Si$\;${\sc iv}~$\lambda$1400, C$\;${\sc iv}~$\lambda$1550, 
and N$\;${\sc iv}~$\lambda$1718) and the iron ``forest". Data obtained in the NUV region 
reveal fewer lines which are mostly in absorption. The most prominent is 
C$\;${\sc iii}~$\lambda$2296, which we present in Figure \ref{nuv}.

\begin{table}
\caption{Log of {\em HST/STIS} observations of HD~191612. Rotational phases are from the ephemeris given 
by Howarth et al. (2007): JD = 2453415.2 $\pm$ 537.6$E$. The phase uncertainty is about 0.002. The 
total exposure time is given in seconds.}
\begin{tabular}{lcccc}
\hline\hline
Grating &  Date (UT) & MJD (mid) & exp. time & Phase $\phi$   \\
\hline
E140M      & \multicolumn{1}{c}{2010-08-23} & \multicolumn{1}{c}{55431.089} &  1500 & 0.751  \\
E230M      & \multicolumn{1}{c}{2010-08-23} & \multicolumn{1}{c}{55431.104} &   300 & 0.751  \\
E140M      & \multicolumn{1}{c}{2011-01-04} & \multicolumn{1}{c}{55565.294} &  1500 & 0.000   \\
E230M      & \multicolumn{1}{c}{2011-01-04} & \multicolumn{1}{c}{55565.310} &   300 & 0.000    \\
E140M      & \multicolumn{1}{c}{2011-05-18} & \multicolumn{1}{c}{55699.291} &  1500 & 0.250   \\
E230M      & \multicolumn{1}{c}{2011-05-18} & \multicolumn{1}{c}{55699.303} &   300 & 0.250   \\
E140M      & \multicolumn{1}{c}{2011-09-30} & \multicolumn{1}{c}{55834.512} &  1500 & 0.501  \\
E230M      & \multicolumn{1}{c}{2011-09-30} & \multicolumn{1}{c}{55834.527} &   300 & 0.501  \\
\hline
\label{obsdetails}
\end{tabular}
\end{table}

There is significant line-profile variability at a level which exceeds
that normally found in O-star spectra\footnote{Many O stars are known 
to exhibit resonance-line variability in the form of discrete absorption 
components (DACs); see, e.g., Kaper et al. (1996). DACs are localized optical 
depth enhancements that propagate through P-Cygni absorption troughs from red to blue 
on a timescale of typically $\sim$days, becoming narrower as they approach the terminal velocity.  
The variations seen here are fundamentally different in character, insofar as the change 
in absorption depth occurs over the entire width of the line profile.}, although the amplitudes are
nonetheless modest compared to the substantial changes reported in the
optical spectrum (see, for example, Fig.~5 of Walborn et al. 2003).
The main changes are observed in the absorption troughs of the
N$\;${\sc v}~$\lambda$1240 and C$\;${\sc iv}~$\lambda$1550 P Cygni profiles, together with
the high-velocity part of N$\;${\sc iv}~$\lambda$1718, which are all stronger at
$\phi = 0.0$ (the `high state') than at other phases, while the
(essentially photospheric) iron-forest absorption weakens at this phase (Figure \ref{uv}).
In contrast, both Si$\;${\sc iv}~$\lambda$1400 and C$\;${\sc iii}~$\lambda$2296 are
weaker at $\phi = 0.0$ (see Figure \ref{nuv}). The C$\;${\sc iii}~$\lambda$1176 
multiplet (not shown) is very weak compared to these 
two features but follows the same trend. This behaviour is related to 
subtle magnetic field effects on the wind and is discussed 
more in detail in Section \ref{mhd}.

The Of?p star HD~108 also exhibits greater wind absorption in C$\;${\sc iv} 
and N$\;${\sc v} and weaker absorption in the iron forest during the high state.
However, there are important differences between the two stars. 
For example, in HD~191612 (unlike HD~108) the C$\;${\sc iv} and
N$\;${\sc v} wind profiles never become saturated, regardless of the phase.
Moreover, in HD~191612 (again, unlike HD~108) the Si$\;${\sc iv}~$\lambda$1400 
behaves in the opposite way to C$\;${\sc iv} and N$\;${\sc v}, the absorption 
being strongest at phase 0.5. Such differences may be related to the different 
stellar and magnetic properties of these stars. Different mass-loss rates, 
magnetic obliquity, and hence viewing angle may all 
affect the observed line profile variations.

Table~\ref{linebehavior} summarizes this discussion of the 
relative variability of the main UV features of HD~108 and HD~191612. 
Note that the analysis of the C$\;${\sc iii} lines in HD~108 is 
hindered by the very low signal-to-noise of the {\em IUE} observations available.

\begin{figure*}       
\centering       
\includegraphics[width=13.5cm,height=17.0cm,angle=90]{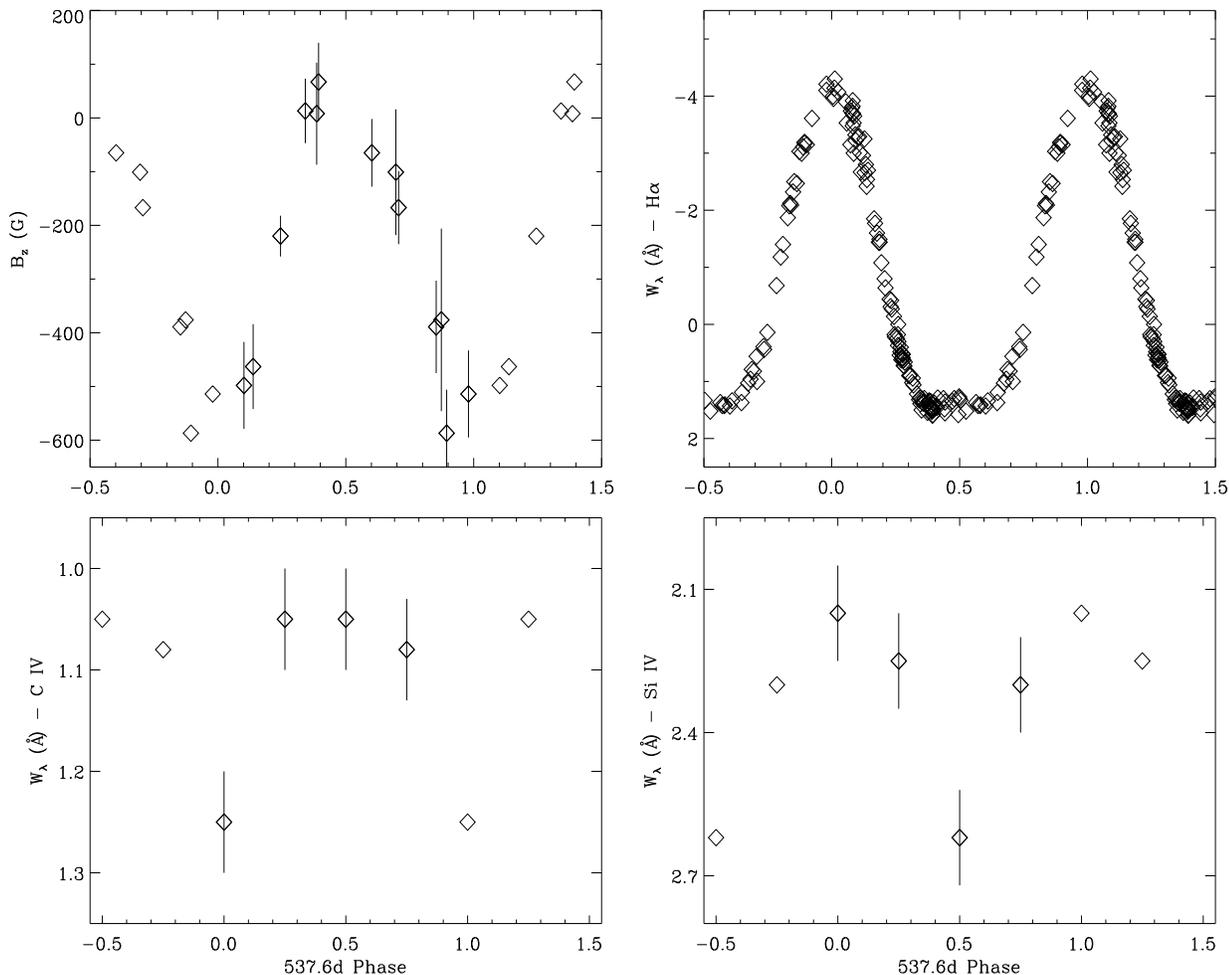}       
\caption{Phase-resolved data of HD~191612. Upper panels: Longitudinal magnetic field component (left) and H$\alpha$ equivalent width 
(right; data from Howarth et al. 2007; Wade et al. 2011). Lower panels: equivalent width of C$\;${\sc iv}~$\lambda$1550 and Si$\;${\sc iv}~$\lambda$1400 obtained from the new {\em STIS} observations.}      
\label{pharesdata}       
\end{figure*}  



\subsection{Equivalent width measurements}
\label{eqw}

To quantify the variability, we measured equivalent widths for the main UV lines, 
using fixed wavelength intervals for each feature (e.g., 1534.0-1551.3\AA\, for C$\;${\sc iv} 
and 1390.2-1395.0\AA\, for Si$\;${\sc iv}). In Fig.~\ref{pharesdata} we present the results for 
C$\;${\sc iv} and Si$\;${\sc iv}, along with H$\alpha$ equivalent 
widths and the longitudinal magnetic-field values, $B_Z$ (from Howarth et al. 2007; Wade et al. 2011). 
Maximum C$\;${\sc iv}~$\lambda$1550 and minimum Si$\;${\sc iv}~$\lambda$1400 absorptions 
occur at $\phi = 0$, which coincides with maximum H$\alpha$ emission. The UV iron-forest 
strength (not shown), as well as other UV profiles (e.g. N$\;${\sc v}~$\lambda$1240 and 
C$\;${\sc iii}~$\lambda$2296), also follows the rotational variability, along with 
photometry from Hipparcos (Walborn et al. 2004). In summary, the UV profiles of HD~191612 vary
in concert with other measurements: \emph{the UV variability follows 
the established 537.6-d period} and results from rotational modulation.

The oblique rotator model was proposed for HD~191612 by
Donati et al. (2006), with details subsequently refined by Howarth et al. (2007) and 
Wade et al. (2011).  The currently accepted model has 
a (rotational) axial inclination of $i \simeq 30^\circ$, a
centred dipole magnetic field with an obliquity of $\beta \simeq
67^\circ$, and a polar strength of $\sim$2.5kG.  
With this geometry, the line of sight is almost aligned with the 
dipole axis at rotational phase $\phi \simeq 0$, and almost 
perpendicular to it at $\phi \simeq 0.5$.
In such a scenario, we expect enhanced density and slower velocities in the
region of the magnetic equator, which is fed by magnetically deflected outflows
from the polar and temperate latitudes (ud-Doula \& Owocki 2002).
Thus, \emph{a priori,} the net effect expected for a wind profile at $\phi = 0.5$
is less absorption at high velocities, as exhibited, e.g., by 
C$\;${\sc iv}~$\lambda$1550 and N$\;${\sc v}~$\lambda$1240. 
At a first glance, the Si$\;${\sc iv}~$\lambda$1400 profile (and C$\;${\sc iv}~$\lambda$2296) 
seems to contradict this idea. We come back to this question later in Section \ref{mhd}, 
where we show that effects from a {\em dynamical magnetosphere} may solve this puzzle.

\begin{table}
  \caption{The high-state ($\phi \simeq 0.0$) behaviour of the main UV features 
    in HD~191612 and HD~108 compared to low-state spectra.
    The +abs ($-$abs) notation indicates greater
    (less) absorption in the photospheric profiles or 
in the absorption part of a P Cygni profile in the high state.  Note
the different behaviours for Si$\;${\sc iv} and C$\;${\sc iii}. }
\begin{tabular}{lcc}
\hline\hline
Feature &  HD 191612    &  HD 108     \\
\hline
N$\;${\sc v}~$\lambda$1240   & +abs        & +abs   \\
C$\;${\sc iv}~$\lambda$1550  & +abs        & +abs   \\
N$\;${\sc iv}~$\lambda$1718  & +abs        & +abs   \\
Fe$\;${\sc iv} forest        & $-$abs      & $-$abs \\
C$\;${\sc iii}~$\lambda$1176 & {\bf -abs}  & {\bf ?} \\
Si$\;${\sc iv}~$\lambda$1400 & {\bf -abs}  & {\bf +abs} \\
C$\;${\sc iii}~$\lambda$2296 & {\bf -abs}  & {\bf ?}  \\
\hline
\label{linebehavior}
\end{tabular}
\end{table}


\section{ATMOSPHERE MODELS}
\label{models}

For the oblique-rotator configuration described above, 
spherical symmetry is clearly a very poor approximation for lines principally
formed inside the Alfv\'en radius; it is precisely this  
lack of spherical symmetry (coupled with fortuitous viewing geometry)
that generates the strong rotational modulation observed in H$\alpha$ and some 
optical helium lines.  This conclusion is strengthened by the
success that Sundqvist et al. (2012) 
had in reproducing the observed variability of H$\alpha$ by using 
the structure of a magnetically confined wind predicted by MHD models.

Sundqvist et al. (2012) coupled their MHD models to relatively 
simple radiative-transfer calculations appropriate to H$\alpha$ line
formation.  However, this work has not yet progressed to the point
where fully self-consistent modelling of all UV lines is possible 
(see however Section \ref{mhd}). In order to obtain a quantitative 
characterization of our {\em STIS} data, we therefore first turn to the 
other extreme of spectral modelling, embodied in the CMFGEN code (Hillier \& Miller 1998). 
This code solves the equation of radiative transfer with considerable attention to the microphysics
(line blanketing, NLTE, radiative and statistical equilibria), but
with simplified -- normally spherical -- symmetry.

While spherical symmetry is obviously a poor assumption for lines
formed in the vicinity of the Alfv\'en radius, we might expect that
features formed in or close to the photosphere, such as the iron
forest, can reasonably be modelled under this approximation.
Furthermore, because the magnetic dipole energy density falls much
faster with radial distance than does the wind kinetic energy density
($\sim{r}^{-6}$ vs. $r^{-2}$), the stellar-wind outflow must become
essentially radial at large distances.  With some caution, we may
therefore investigate lines formed at large radii --
such as the high-velocity parts of UV P Cygni profiles -- under the
simplifying assumption of spherical symmetry. 

In order to further avoid non-spherical effects, we focused on the analysis 
of {\em STIS} data at phase $\phi = 0$, when the magnetic pole is oriented 
towards the observer (see Fig. \ref{pharesdata}). At this configuration, 
the line-of-sight absorption column will consist mostly of wind plasma streaming 
from the open field lines above the magnetic pole, which presumably then leads 
to a reduced influence of the dense, confined wind material concentrated near 
the magnetic equator. Thus, for UV wind lines, a spherically symmetric model 
should be a better approximation in the high state.

\begin{figure*}                          
\includegraphics[width=11cm,height=17cm,angle=90]{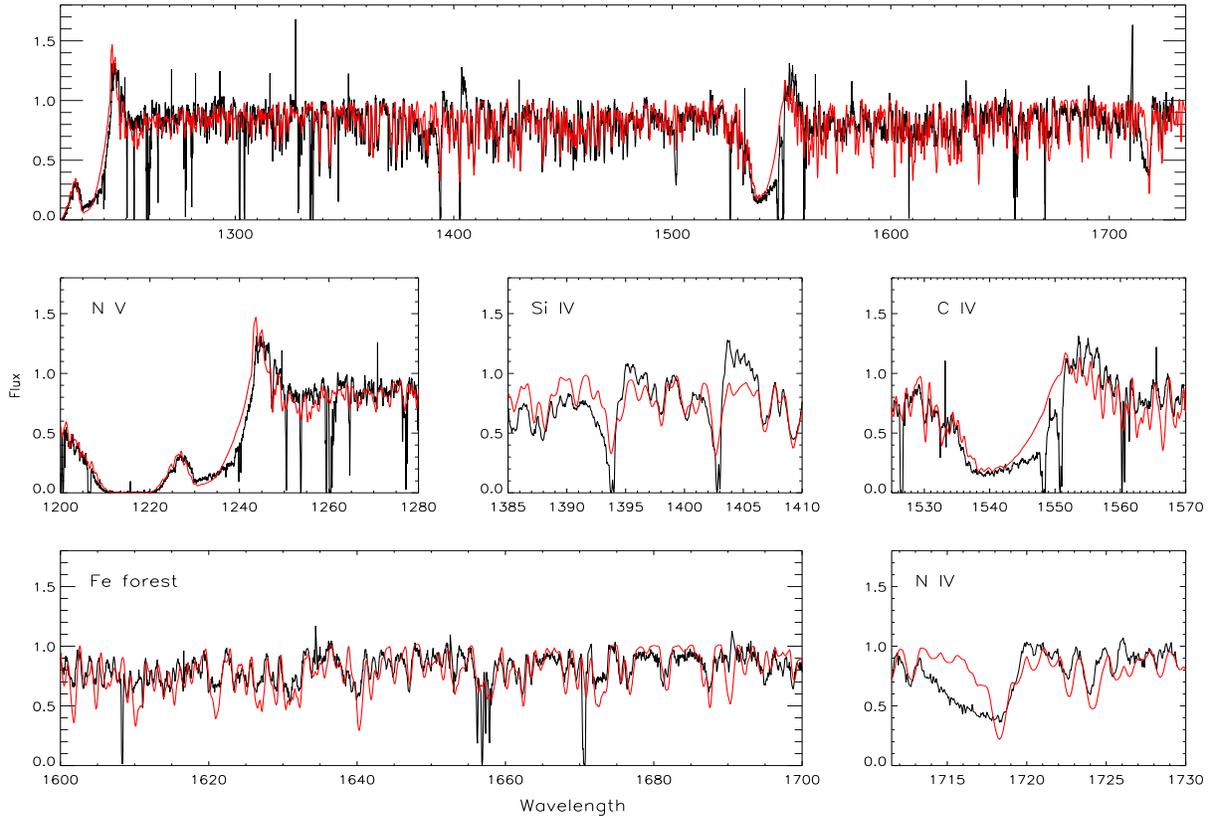}       
\caption{Our adopted CMFGEN model (red) compared to {\em STIS} data of HD~191612 at 
phase $\phi = 0.0$ (black). Wavelengths are given in angstroms and the flux is 
normalized. H$\;${\sc i} interstellar absorption is accounted for shortward the N$\;${\sc v} profile; log N(H$\;${\sc i}) = 21.48. }      
\label{uvfit}       
\end{figure*}

In Table \ref{params} we summarize the stellar and wind properties inferred. Our final model 
is compared in Figure \ref{uvfit} with normalized {\em STIS} spectra. The effects of clumping were accommodated
with a filling-factor formalism ($f_{cl} = 0.1$), beginning at an expansion velocity of 30 km s$^{-1}$ ($v_{cl}$); 
we also allowed for the the effects of X-rays produced by wind instabilities, adopting $\log{L_X/L_{\rm Bol}} = -6.1 \pm 0.1$ (Naz\'e et al. 2010).

Below we discuss other parameters and the issues found during the analysis.

\begin{figure*}       
\centering       
\includegraphics[width=8cm,height=17.5cm,angle=90]{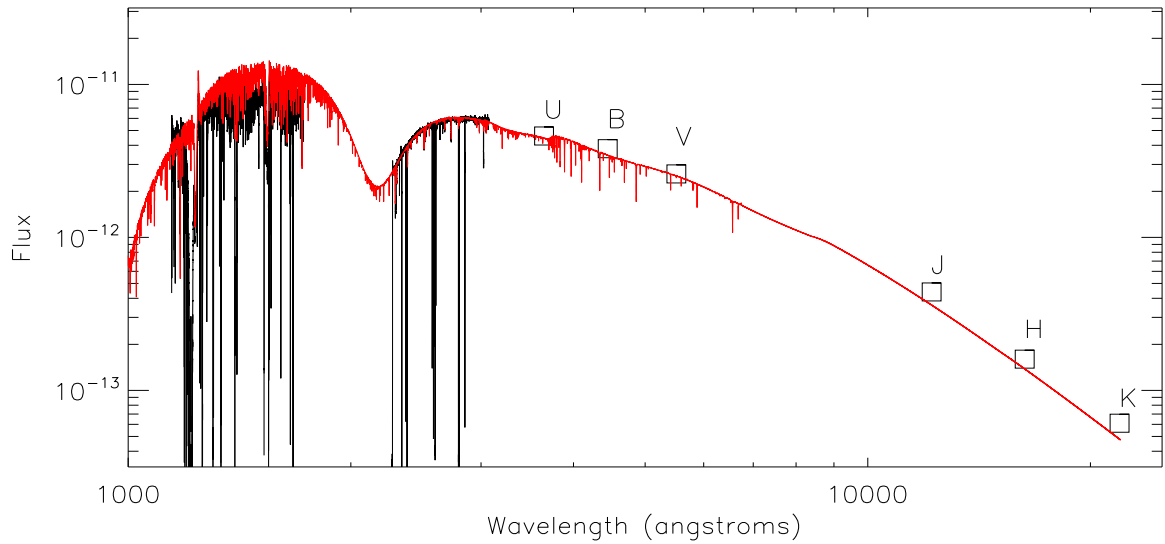}    
\caption{{\em STIS} FUV and NUV data plus UBVJHK photometry converted to flux points (black) and our adopted CMFGEN model (red). The adopted 
distance is 2.3 kpc. Flux units are erg cm$^{-2}$ s$^{-1}$ \AA$^{-1}$.}      
\label{sed}       
\end{figure*}  


\subsection{Chemical composition, rotation and macroturbulence}

Following Howarth et al. (2007), we adopted a macroturbulent velocity $v_{mac} = 45$ km s$^{-1}$ and projected 
equatorial rotation velocity $v_e {\rm sin\,} i$ of zero. We note, however, that the UV observations are equally 
well matched by any model having

\[ 
\sqrt{v_{mac}^2 + \left({v_e {\rm sin\,} i}/{2}\right)^2} \simeq 45 \textnormal{km}\,s^{-1}. 
\]

We used the solar chemical composition (Grevesse et al. 2010), except for nitrogen.  
The nitrogen content in HD~191612 was recently determined by Martins et al. (2011) based 
on the analysis of photospheric lines in high-resolution optical data. 
Here, we use their results, which fix N/H at $2.7 \times 10^{-4}$ (the solar value is $6.8 \times 10^{-5}$; 
by number).


\begin{table}
\caption{Revised stellar and wind properties of HD~191612 (CMFGEN models).}
\begin{tabular}{ll}
\hline
T $_{eff}$ (K)           & 36000 $\pm$ 2000 \\
log $g$ (cgs)            & 3.75 $\pm$ 0.20     \\
log L$_\star$/L$_\odot$   & 5.45 $\pm$ 0.10    \\
R$_{\star}$/R$_\odot$     & 13.7 $^{+2.3}_{-2.0}$ \\
M$_{\star}$/M$_{\odot}$    & 39  $^{+28}_{-17}$    \\
$v_{\rm eq}\sin{i}$   (km s$^{-1}$) & 0  \\
$v_{mac}$ (km s$^{-1}$)   & 45  \\
$\xi _t^{phot}$  (km s$^{-1}$)  & 15  \\
P$_{rot}$ (days)         & 537.6  \\
\hline
log $\dot{M}$           & -7.9 $\pm$ 0.3  \\
v$_\infty$ (km s$^{-1}$)    & 2400  $\pm$ 500 \\
f$_{cl}$                  & 0.1 \\
v$_{cl}$ (km s$^{-1}$)     & 30 \\
log $L_X/L_{BOL}$        & -6.1  \\
distance (pc)           & 2300 \\    
\hline
\end{tabular}
\label{params}
\end{table}


\subsection{Luminosity and distance}

HD~191612 is probably a member of the Cyg OB3 association, at a
distance of 2.3 kpc (Humphreys 1978).  Adopting this distance, we estimate the
luminosity from a fit to the FUV and NUV spectra plus
UBVJHK magnitudes available in the Galactic O-Star Catalog
(Sota et al. 2008).  The results are included in Table~\ref{params}, and the fit
displayed in Fig.~\ref{sed}.  The differential extinction $E(B-V)$ was estimated 
from the observed $B-V =0.26$ and the typical intrinsic value for O6--O8 stars, $(B-V)_0 =
-0.27$ (Martins \& Plez 2006).

We note that HD~191612 is known to be photometrically
variable, but the amplitude is too small to affect these results significantly;  e.g., the
$H_p$ magnitude varies by not more than $\sim$0.05.  We also neglect the contribution of the 
$\sim$early-B secondary. Such a star is much cooler ($\sim$20kK) and fainter than the
primary and is expected to contribute only about 10 per cent in the visible, and 
considerably less at UV wavelengths (see Howarth et al. 2007).


\subsection{Issues}

From Fig.~\ref{uvfit} we see that the agreement between our atmosphere model 
and the {\em STIS} data of HD~191612 at phase $\phi = 0$ is reasonable. We predict P Cygni 
profiles for N$\;${\sc v}~$\lambda$1240 and C$\;${\sc iv}~$\lambda$1550, a 
good fit for the iron forest, and absorption/incipient P Cygni profiles to Si$\;${\sc iv}~$\lambda$1400 and 
N$\;${\sc iv}~$\lambda$1718, as observed. However, it is evident that some features  
could not be reproduced in detail. For example, the synthetic N$\;${\sc iv}~$\lambda$1718 profile 
lacks blueshifted absorption. By increasing the N/H ratio to $\sim 8 \times 10^{-4}$ we
improve the fit to the observed profile, but this value is much higher
than that reported by Martins et al. (2012), even considering the 
uncertainties in the determination.  
We have also tried several ways to improve the fit to the Si$\;${\sc iv}~$\lambda$1400 
feature (see Fig.~\ref{uvfit}), although a comparison is handicapped by interstellar absorption.

Some discrepancies were already expected in our analysis, since we used 
spherically symmetric atmosphere models. As discussed previously, 
lines formed closer to the Alfv\'en radius may be significantly affected by 
non-sphericity. Ideally, MHD wind-confinement simulations coupled with detailed 3D 
radiative-transfer calculations should be used to derive the 
stellar and wind properties of HD~191612. However, given the complexity of 
such a task, the calculation of a full UV and optical theoretical 
spectrum -- as is done by 1D models -- is not currently feasible.
Instead, in the next section we attempt to provide a  
physical interpretation for generic UV profiles based on MHD simulations.


\section{UV WIND LINES CALCULATED FROM MHD MODELS}
\label{mhd}

This section reports initial results from ongoing attempts to use
radiation MHD wind simulations (developed first
by ud-Doula \& Owocki 2002) to analyze the strengths and variability
of UV resonance lines in magnetic massive stars. To this end, we apply 
the same MHD simulation used previously to model the rotational phase
variation of HD\,191612's H$\alpha$ line (Sundqvist et
al. 2012), however modifying the radiative transfer code developed there 
to synthesize UV scattering lines (following the `3-D SEI' method,
e.g. Cranmer \& Owocki 1996).

\begin{figure}
\resizebox{\hsize}{!}{\includegraphics[angle=90]{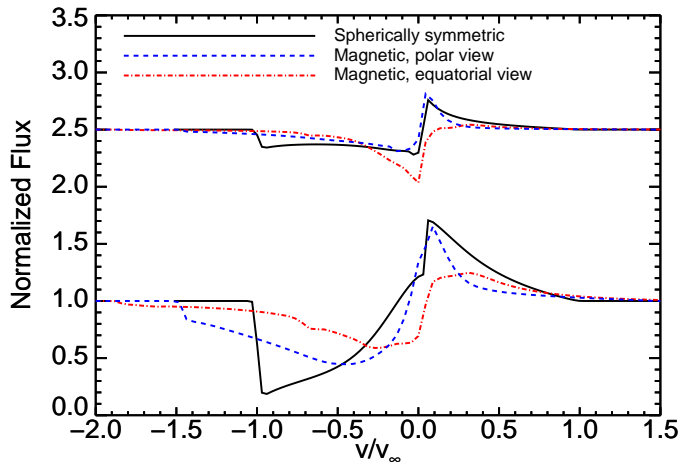}}
\caption{Synthetic UV scattering lines computed from MHD wind simulations for an observer  
viewing the star from above the magnetic pole (blue, dashed lines) and equator (red, dashed-dotted lines),
as compared to profiles computed from a spherically symmetric, non-magnetic wind model (black lines).
The lower (upper; shifted by +1.5) panel shows profiles computed using a constant line-strength parameter 
$\kappa_0 = 1$ ($\kappa_0 = 0.1$), where $\kappa_0$ is proportional to the wind mass-loss rate and 
the ion fraction (Puls, Vink, \& Najarro 2008). The abscissa shows the wind velocity normalized to the 
terminal speed of the non-magnetic model.}
\label{Fig:mag_uv}
\end{figure}

Fig.~\ref{Fig:mag_uv} shows generic synthetic profiles for a
weak (upper panel) and moderately strong (lower panel) line, 
calculated i) from the MHD model for an observer viewing 
from above the magnetic pole (``high state'') and
equator (``low state''), and ii) from the spherically symmetric wind
model used as initial condition to the MHD simulation. Note how the
phase trends of the modeled absorption-strengths indeed display opposite
behaviour for the weak and strong lines, as observed for the (weaker)
Si$\;${\sc iv} and (stronger) C$\;${\sc iv} lines in HD\,191612 (see Fig. \ref{uv}).

This anti-correlated variability of strong and weak lines can be qualitatively 
understood as follows: the line-of-sight absorption column in the low state samples the
low-velocity, high density wind material trapped by closed field lines
near the magnetic equator in HD\,191612's \textit{dynamical
  magnetosphere} (Sundqvist et al. 2012). By contrast, the absorption
column in the high state covers a wider range of velocities and
densities, since it stems from the open field lines streaming from the
magnetic pole, and so more closely resembles a `normal' wind outflow.
The result is that a \textit{strong} line,
which has sufficient optical depth to produce measurable absorption at 
high velocities, is stronger in 
the \textit{high state} (lower panel). On the other hand, a
\textit{weak} line, which is not able to produce significant absorption at high velocities, is
stronger in the \textit{low state} (upper panel), because during that 
phase the line can absorb from the high density, low velocity material 
trapped in the wind's dynamical magnetosphere.

This figure also shows that profiles computed for intrinsically strong lines 
(e.g., those corresponding to the line strength parameter $\kappa_0 = 1$ in 
Fig.~\ref{Fig:mag_uv}) are systematically weaker than their counterparts computed from spherically symmetric models,
particularly at high velocities.

This effect can be understood by considering 
the faster-than-radial expansion of a radial polar flow-tube in a magnetic 
wind (Kopp \& Holzer 1976; Owocki \& ud-Doula 2004), which leads to 
a different density profile as function of velocity than in a spherical wind. Following 
Owocki \& ud-Doula (2004), let us parameterize 
the non-radial expansion factor of an areal element, $A(r')$:

\[  h(r') \equiv \frac{A(r')}{r'^2} \approx  \frac{r'(1+\eta_\star^{3/8})}{r'+\eta_\star^{3/8}} \]

\noindent
where $r' \equiv r/R_\star$ is the scaled radius and $\eta_\star$ is the 
magnetic wind confinement parameter applicable to a dipolar field 
(ud-Doula \& Owocki 2002). Note how $h$ starts off at unity, but increases with radius reaching the asymptotic value 
$h(r' \rightarrow \infty) = 1+\eta_\star^{3/8}$. By the equation of continuity $\rho \sim 1/(Av)$;
consequently, for equal basal mass fluxes, the density $\rho$ of a non-magnetic wind will be higher
at a specified  velocity in the outer wind than the density $\rho_{\rm B}$ of a magnetic one, by a factor 
$\rho/\rho_{\rm B} \approx (1+\eta_\star^{3/8})$. Since $\eta_\star \approx 50$ for HD\,191612, 
$\rho/\rho_{\rm B} \approx 5.3$. This effect may partly explain the low mass-loss rate inferred from 
CMFGEN models (Sect. \ref{models}; dilution effect neglected) compared to the rate needed in the MHD simulations 
to reproduce HD~191612's observed H$\alpha$ emission strength and variability  
($\sim 10^{-6}$ M$_\odot$ yr$^{-1}$). Note also from Fig.~\ref{Fig:mag_uv} that the MHD simulation further predicts a significant 
increase in wind terminal speed, here by $\sim 50 \%$. However, this does not seem to be 
observed for HD\,191612, whose $v_\infty \approx 2400 \rm \, km/s$ is 
not much higher than typical values inferred for non-magnetic O-stars.

A detailed study of the formation of UV wind lines in magnetic
massive stars, focusing on issues like those discussed here,
will be reported separately (Sundqvist \& ud-Doula, in prep.). 
 
\section{SUMMARY AND CONCLUSIONS}
\label{concs}

We have presented for the first time phase-resolved UV data for an Of?p
star, HD~191612. Our main findings and conclusions are outlined below.

\begin{itemize}

\item The spectra observed at different rotational phases are broadly
  similar, with the same set of wind and photospheric lines consistently
  present. However, line profile variations are very significant  
  compared to the ones found in {\em normal} O stars.  

\item The variability observed is not as drastic as that reported for
  H$\alpha$ and the He I lines in the optical part of the spectrum. We
  interpret this result as indicative of the stronger influence of the
  magnetic field on the lines formed closer to the photosphere (i.e.,
  inside the Alfv\'en radius), where we find confined material.

\item We observe more wind absorption in N$\;${\sc v}~$\lambda$1240 and 
  C$\;${\sc iv}~$\lambda$1550 and a less intense iron-forest spectrum at $\phi = 0$ 
  compared to $\phi = 0.5$. The former phase corresponds approximately to the 
  configuration where the magnetic pole is towards the observer while the 
  latter corresponds to the line of sight lying in the plane of the magnetic 
  equator. Such variability is very similar to that reported for HD~108 in Paper~I, 
  and could be considered as probably representative of Of?p stars. There
  are, however, notable differences between HD~108 and HD~191612, such
  as the absence of saturated P Cygni profiles in HD~191612 and the
  anti-correlated variability of Si$\;${\sc iv}~$\lambda$1400 and other wind lines.

\item CMFGEN models are capable of providing a reasonable fit to 
  the UV spectrum of HD~191612 at phase $\phi = 0$. The wind and photospheric lines are 
  correctly predicted in broad terms, but not all profiles are 
  reproduced in detail. Furthermore, spherically symmetric models 
  cannot reproduce the variability revealed by the {\em STIS} data, since 
  these variations arise from the dynamical magnetosphere of HD~191612. 
 

 \item We use MHD simulations in combination with simplified radiative transfer calculations 
 to analyse the new {\em STIS} phase resolved data of HD~191612. We compare generic UV profiles 
 representing C$\;${\sc iv}~$\lambda$1550 and Si$\;${\sc iv}~$\lambda$1400. The observed behaviour 
 at phases $\phi = 0.0$ and $\phi = 0.5$ are qualitatively reproduced for both lines; an explanation for the 
 anti-correlated variability (Figs. \ref {uv} and \ref{pharesdata}) is provided in terms of line strengths.

 \item The mass-loss rate derived for HD~191612 from non-LTE expanding atmosphere models for 
 the UV is considerably lower than the one inferred from the MHD simulations (based on H$\alpha$; 
 $\sim 10^{-6}$ M$_\odot$ yr$^{-1}$). This discrepancy can be partially understood by a density dilution 
 effect (Sect. \ref{mhd}) but deserves a deeper investigation.

\end{itemize}

The present results take us a further step towards a detailed picture
of the optical and UV variability in Of?p stars.  In general, the
optical spectra are strongly affected by the magnetic field, with line
formation at or below the Alfv\'en radius. The UV P Cygni profiles
present moderate changes, suggesting spherical symmetry may be a
reasonable first approximation for the wind at large radii. 
MHD simulations such as the ones computed here are however mandatory to describe the 
(phase-resolved) UV profiles in detail.

Two additional Of?p stars have been recently detected to be magnetic, namely, NGC 1624-2 and 
CPD -28 2561 (Walborn et al. 2010; Wade et al. 2012; Hubrig et al. 2012). The basic results regarding the ultraviolet 
spectrum reported here and in our previous paper are expected to be also observable in these stars.
However, for NGC 1624-2, future UV observations will likely reveal a considerably larger 
variability in the wind profiles than the ones seen so far, since its dipole field is estimated to be 
about 20kG (Wade et al. 2012). Such a field is about 8 times more intense than in any other known magnetic O-type star. 

The Of?p class now stands out among the O-type stars as magnetic variables. They 
represent therefore an ideal astrophysical laboratory for the study of stellar wind confinement and 
magnetic field effects in the evolution of massive stars. The progress achieved so far shows that 
multiwavelength analyses are crucial for a complete description of their properties.

\section*{Acknowledgments}
WLFM acknowledges support from the Funda\c c\~ao de Amparo 
\`a Pesquisa do Estado do Rio de Janeiro (FAPERJ/APQ1). JCB 
was supported by NASA grant NNX08AC146 to the University of 
Colorado at Boulder during the completion of this work. 
JOS acknowledges support from NASA ATP grant NNX11AC40G. 
NRW acknowledges support provided by NASA through grant  
GO-12179.01 from STScI, which is operated 
by AURA, Inc., under NASA contract NAS5-26555.
GAW acknowledges Discovery Grant support from the Natural Sciences and Engineering 
Research Council of Canada (NSERC). AuD acknowledges support from NASA ATP grant NNX12AC72G. 
We thank the French Agence Nationale de la Recherche (ANR) for financial support.

{}

\label{lastpage}
\end{document}